\documentclass[referee]{aa} 

\usepackage{graphicx}
\usepackage{txfonts}
\begin{document}

   \title{Search for torsional oscillations in isolated sunspots}


   \author{A. B. Griñón-Marín\inst{1,2}, 
          H. Socas-Navarro\inst{1,2},
          \and
          R. Centeno\inst{3}
          }

   \institute{Instituto de Astrof\'{\i}sica de Canarias,
              V\'{\i}a L\'actea, 38205 La Laguna, Tenerife, Spain\\
         \and
              Universidad de La Laguna, Departamento de Astrof\'{\i}sica, 
              38206 La Laguna, Tenerife, Spain \\
         \and
             High Altitude Observatory (NCAR),
             3080 Center Green Dr. Boulder CO 80301 \\
             }


 
  \abstract
   {In this work we seek evidence for global torsional oscillations in alpha sunspots. We have used long time series of continuum intensity and magnetic field vector maps from the Helioseismic and Magnetic Imager (HMI) instrument on board the Solar Dynamics Observatory (SDO) spacecraft. The time series analysed here span the total disk passage of 25 isolated sunspots. We found no evidence of global long-term periodic oscillations in the azimuthal angle of the sunspot magnetic field within $\sim$ 1 degree. This study could help us to understand the sunspot dynamics and its internal structure.}


   \titlerunning{Search for torsional oscillations in isolated sunspots}
   \authorrunning{Gri\~n\'on-Mar\'in et al. 2017}
   \maketitle

%

\section{Introduction}
The issue of long-term (on scales of several hours to days) morphological changes in sunspots, and particularly the possible existence of apparent rotational motions and oscillations, has drawn attention since the early 20th century.  Relevant previous observations include photoheliograms (\citealt{M71}, \citealt{K82}, \citealt{KUC82}, \citealt{A83}, \citealt{P85}, \citealt{GL87}, and \citealt{GLK88}) and in some cases radial velocity measurements (\citealt{E1909}, \citealt{K52}, \citealt{A57}, \citealt{K76}, \citealt{G77}, \citealt{G81}, \citealt{G85}, and \citealt{G86}) using ground-based telescopes. Obviously, these observations are very difficult to conduct from the ground owing to the limited time coverage (weather, day-night cycle, seeing, etc). 

Torsional oscillations are attributed to tensile forces of the magnetic field lines, which tend to drive a perturbed system back to its unperturbed lower energy state. When an external force acts on the field line twisting it in one direction, this tensile restorative force kicks in, counteracting the displacement and driving the system back to its original state. However, excess kinetic energy makes the system overshoot into the opposite twist, generating an oscillatory motion. The detection and characterization of torsional oscillations would be very interesting, as they would provide a valuable test for MHD models. Whenever a physical system exhibits proper oscillation modes, they become a useful tool to explore the structure of such a system (\citealt{KC15}). 

The advent of the Solar and Heliospheric Observatory (SOHO; \citealt{DFP95a}, \citealt{DFP95b}, \citealt{DFP95c}, and \citealt{D95}) permitted, for the first time, observations with continuous time coverage but limited to imaging, line-of-sight magnetograms, and a spatial resolution of 1\arcsec.2 \ around the disk centre only (outside this region, the 4\arcsec resolution was too coarse for torsional oscillation studies). In practice, the permitted time coverage was then restricted to only a few days. Some of the resulting observational papers on torsional oscillations are those of \citealt{HD03}, \citealt{N04}, \citealt{GG05}, \citealt{GG06}, and \citealt{GK11}. Torsional oscillations, from the theoretical point of view, have been studied to obtain some properties of the interior of the sunspots, for example the length of the magnetic tube of the sunspot (\citealt{G84}, \citealt{S84}, \citealt{N97}). By assimilating observations into their models these authors were able to obtain physical information about depths that are inaccessible to direct observations.

The Solar Dynamics Observatory (SDO; \citealt{PTC12}) space mission has opened a new window for unprecedented long-term studies of active region evolution. Furthermore, the 11-year solar activity cycle peaked in 2014 and sunspots have been abundant on the solar disk for a few years now. The Helioseismic and Magnetic Imager (HMI; \citealt{SSB+2012}, \citealt{SSB+12}) on board the SDO routinely measures the full magnetic field vector over the full solar disk, which allows us to track it with consistent image quality. It is the ideal instrument to analyse the evolution of sunspots, and in particular the azimuthal component of the penumbral magnetic field with good time coverage.

Some previous works on torsional oscillations have reported periods of several days (\citealt{G81}, \citealt{G85}, \citealt{P85}, \citealt{GL87}, \citealt{HD03}, \citealt{N04}, \citealt{GG05}, \citealt{G10}, and \citealt{GK11}). Table \ref{tab-previous_results} shows the periods and amplitudes of the oscillations reported in these works, as well as the length of the time sequence used in each investigation. In all of these cases, the observations lacked the time span and resolution necessary to support with confidence the oscillation periods claimed by the authors. This is not surprising, given the limitations imposed by ground-based observations, especially regarding time coverage and cadence. The SDO/HMI opens new opportunities to analyse these oscillations with unprecedented detail and time coverage, allowing us, for the first time, to unambiguously detect and characterize sunspot torsional oscillations, if they indeed exist. The presence of this type of oscillations might potentially shed some light on the understanding of sunspot dynamics and could allow us to obtain information on their subsurface structure (\citealt{KC15}). With the good time coverage of the HMI data, we could obtain torsional oscillation period values more accurately than in previous studies, and then determine the sunspot depth with more precision (\citealt{G84}, \citealt{S84}, and \citealt{N97}).

In this work, we seek evidence for oscillations with periods between 24 minutes and 6 days, which is the range that may be reliably probed with the available data. 

\begin{table*}
  \caption{Previous results about torsional oscillations with period values, amplitude of the oscillations, and temporal coverage of each data.}
  \label{tab-previous_results}
  \centering
    \begin{tabular} {l l l l} 
      \hline\hline
      Paper & Period & Amplitude  & Time Sequence \\
       & \lbrack days, except *=min\rbrack & \lbrack deg\rbrack  & \lbrack days\rbrack \\
      \hline\hline
      \citealt{G81} & 6 & - & 4 \\
      \citealt{G85} & 40 * & - & 3  \\
      \citealt{P85} & 7.1 & 40  &  \\
      \citealt{GL87} & 2 - 26 & 4 - 68 & 5 - 8 \\
      \citealt{HD03} & 3 - 5.2 & - & 2.5 \\
      \citealt{GG05} & 3.4 - 7.7 & - & 5 - 8 \\
      \citealt{G10} & 3.3 - 7.7 & - & 5 - 8 \\
      \citealt{GK11} & 3 - 5.2 & - & 2 \\
      \hline
    \end{tabular}
\end{table*}


\section{Data}
\label{observations}
The SDO/HMI vector field pipeline carries out the spectral line inversion of the Stokes profiles of the \ion{Fe}{I} 6173 \AA\ line on the full solar disk using the  Very Fast Inversion of the Stokes Vector code (VFISV; \citealt{BTK+11}, \citealt{CSH+14}). It also selects active region patches on which the magnetic field disambiguation (\citealt{BLC12}) and other higher level data products are computed (\citealt{BSH+14}).

We carried out our analysis over 25 isolated sunspots (Table \ref{tab-sunspots}). In order to select the sample studied, we looked for sunspots that did not significantly change in shape and size (as seen in continuum intensity maps) during their passage across the visible disk. In addition, we discarded sunspots in their forming and decaying phases during the disk passage.

We chose the isolated sunspot of the AR11084 to illustrate our method. This sunspot was followed from June 26 2010, 21:00:00 UT, to July 8 2010, 00:00:00 UT. The total time coverage was 12 days and 3 hours with a time cadence of 12 minutes. The polarity of the sunspot is negative whereas the positive polarity of the active region appears in the form of a surrounding plage. This is a mature sunspot and its size does not change significantly during its passage.

\begin{figure}
  \begin{center}
  \includegraphics[width=0.5\textwidth]{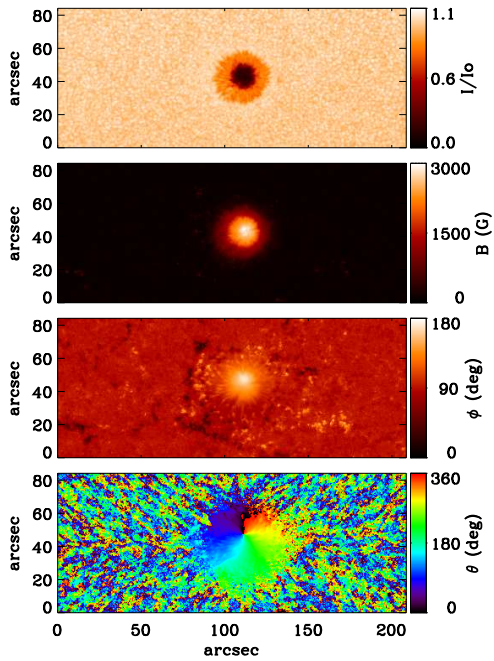}
  \caption{From top to bottom: continuum intensity, magnetic field strength, inclination, and azimuth maps in line-of-sight coordinates when the sunspot of NOAA~11084 was at disk centre.}
  \label{fig-maps}
  \end{center}
\end{figure}

For this work we use a small field of view (58\arcsec x 53\arcsec of size) and we restrict our analysis to four quantities: the continuum intensity, I/I$_o$, the magnetic field strength, B, its inclination with respect to the LOS (line-of-sight), $\gamma$, and its azimuth in the transverse plane, $\psi$ (Fig. \ref{fig-maps}) obtained from the spectral line inversion of the photospheric \ion{Fe}{I} 6173 \AA\ line. We focus the study on the magnetic field component projected on the solar surface. A suitable coordinate transformation from LOS to local solar coordinates and a spherical deprojection are thus required. The procedure is described in the next section.

\section{Methodology}
\label{methodology}
The data provided by the SDO team are given in the LOS reference frame. However, for our purposes the magnetic field vector needs to be expressed in the local, solar reference frame (LOC). Furthermore, since our study involves extremely long time series, perspective effects distort the appearance and geometry of a target as it moves from the East limb to the West one across the solar disk. In order to have a dataset spanning the longest possible duration, we need to de-stretch the field of view to correct for perspective effects, convert the magnetic field coordinates from LOS to LOC and, finally, align the resulting images so that the sunspot does not drift across the field of view. The goal is to produce a homogeneous dataset spanning most of the sunspot passage across the solar disk with a constant perspective, as if we were ``freezing'' solar rotation and observing straight down on the region. Ideally, the only noticeable changes from frame to frame would be due to solar evolution. Obviously, there is some unavoidable loss of information in the process: spatial resolution is degraded when the spot moves away from disk centre due to perspective effects because the horizontal area per pixel increases with the secant of the LOS angle. Moreover, radial structures are seen side-on near the limb. Finally, the line of sight traverses a longer portion of the solar atmosphere when the spot is near the limb, resulting in a slightly higher spectral line formation height. These effects introduce spurious, non-solar changes as the spot moves from the East to the West limb. We restrict our analysis to oscillations of physical parameters that are not systematically affected by such spurious effects, and oscillation periods shorter than half of the disk crossing time.

Three successive steps were necessary in order to prepare the data for this study. First, we changed the reference frame for the magnetic field vector from LOS to LOC. Second, projection effects were removed by applying a geometrical de-stretching of the field of view; third, a local correlation tracking algorithm was developed to align the images. In the following sections, we provide details about each one of the steps.

\subsection{Coordinate change}
The standard data products of the HMI vector field pipeline are the three spherical components of the magnetic field in LOS coordinates: $B_{los}$ (magnetic field strength), $\gamma_{los}$ (inclination), and $\psi_{los}$ (azimuth). For this study, we require the LOC magnitudes, $B_{loc}$, $\gamma_{loc}$, and $\psi_{loc}$, where the inclination is measured with respect to the local solar vertical  (i.e. the radial direction).

\begin{figure}
  \begin{center}
  \includegraphics[width=0.3\textwidth]{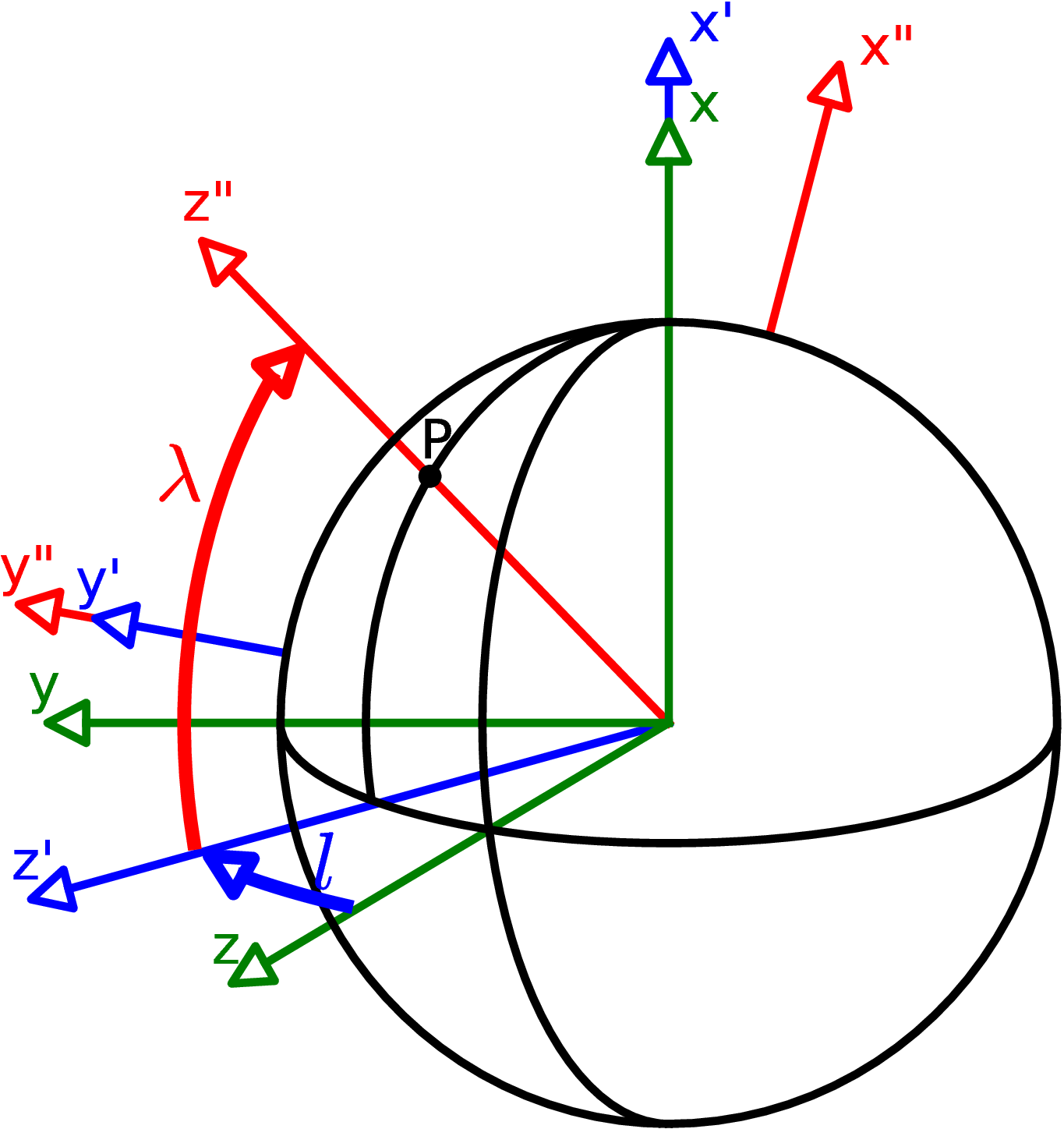}
  \caption{Sketch of the coordinate change from the LOS to the LOC reference frame. The green system represents the LOS (initial) reference frame, the blue one is the reference frame obtained after applying the first rotation around the longitude angle (l), and the red system is the LOC reference frame that is obtained after the second rotation around the latitude angle ($\lambda$).}
  \label{fig-coor}
  \end{center}
\end{figure}

To carry out the transformation, two rotations are performed (see Fig. \ref{fig-coor}). In the first rotation the system is rotated around the $x$-axis, the apparent longitude. In the second rotation, the system is rotated the apparent latitude around the $y$-axis. With this method we ensure that the horizontal direction (y") is parallel to the equator and that z" points northwards.

\subsection{Deprojection}
The second step in the data preprocessing is to apply a geometric deprojection to the images. We re-interpolate the images to a new grid in which the spatial sampling of a pixel remains constant as the active region moves from one limb to the opposite. For each pixel in the field of view we have the Cartesian coordinates $x$ and $y$. Using the solar radius, $R_{\sun}$, we calculate the latitude ($\lambda$) and longitude $(l$) for each pixel, which correspond to the apparent latitude and longitude coordinates, respectively. Now, we can define relative coordinates with respect to a reference position in the field of view $\lambda_{ref}$, $l_{ref}$.

\begin{equation}
  \begin{array}{l}
    \lambda' = \lambda - \lambda_{ref} \, , \\ 
    \textit{l}' = \textit{l} - \textit{l}_{ref} \, , \\ 
  \end{array}
  \label{eq-trans}
\end{equation}

Finally, a linear interpolation is carried out to the new coordinates

\begin{equation}
  \begin{array}{l}
    x' = R_\sun \cos(\lambda') \sin(\textit{l}') \, , \\
    y' = R_\sun \sin(\lambda') \, , \\ 
  \end{array}
  \label{eq-xprima}
\end{equation}
resulting in a set of de-stretched images, corrected for geometrical projection effects.

\subsection{Correlation tracking}
In order to conduct our analysis, we needed a suitable method to track the spot as it travelled across the solar disk, thus ensuring that a given pixel sampled the same spatial position on the Sun throughout the entire time series. 

Correlation tracking is a technique in which a series of images are aligned by finding the optimal shift in the $x$ and $y$ directions to produce the best match of the image with the previous one (or, more generally, a reference frame). In our case, the problem is constrained by the following three facts: (a) The field of view evolves significantly over time. We are tracking an active region for approximately two weeks (the entire solar disk passage), which is comparable to a sunspot lifetime. Furthermore, perspective and projection effects distort the appearance of the entire region at different times in the series. Even if we assumed that the deprojection method works perfectly, spatial resolution and feature contrast change considerably because of the viewing angle; (b) We work with large series of frames, typically composed of thousands of images; and (c) We require very accurate tracking to follow the evolution of the magnetic field at a given location. Some of the analyses presented below are based on averages over large areas of the spot and therefore are not critically sensitive to pixel-scale tracking errors. However, some other analyses are based on the magnetic field evolution at individual spatial locations and for those we require frame-to-frame alignment better than one pixel.

The method followed for the alignment of the time sequence was as follows:

\begin{enumerate}[1]
     \item We enlarged each continuum intensity map to 10 times its original size, interpolating the data linearly to a new grid that was 10 times finer, to obtain subpixel precision.
    \item We calculated the centre of mass of the umbra for each continuum intensity time step.
    \item We compared the centre of mass position to the average of the previous 10 frames. The averaging is performed to reduce the accumulation of random round-off errors that would otherwise produce a slow long-term drift of the image.
    \item We applied the appropriate corrections (i.e. shifts) to each map once we estimated the $x$ and $y$ offsets for each time step in the whole sequence.
\end{enumerate} 

\begin{figure}
  \begin{center}
  \includegraphics[width=0.5\textwidth]{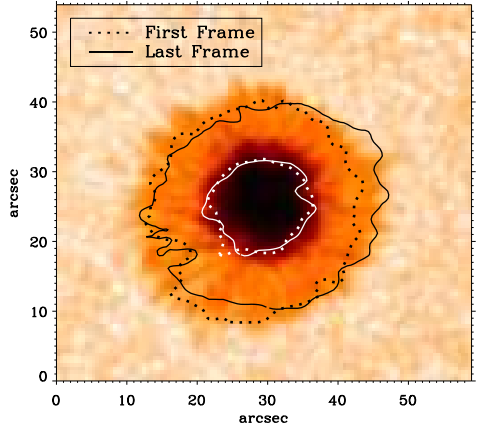}
  \caption{Background image shows the continuum intensity map when the sunspot was crossing the central meridian. The contours show the positions of the umbra and penumbra for the first and last frames of the sequence.}
  \label{fig-align}
  \end{center}
\end{figure}

The goodness of the alignment can be seen in Fig. \ref{fig-align}, where we show the image corresponding to the central frame of the time sequence and overplotted the contours of the umbra and penumbra, six days earlier and six days later, respectively. 

\section{Analysis}
\label{results}
\begin{figure}
  \begin{center}
  \includegraphics[width=0.5\textwidth]{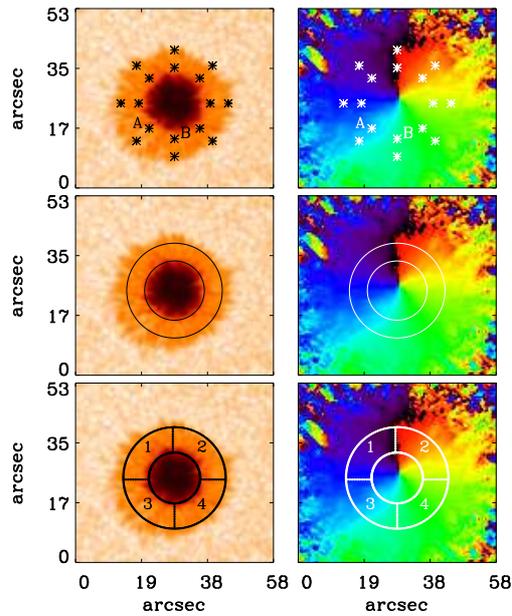}
  \caption{Spatial locations where azimuth variations are analysed in detail. The background images show the continuum intensity (left) and the azimuth (right). Top row: 16 individual pixels, marked by the asterisks, were examined. Middle row: Averages along two concentric circumferences. Bottom row: Averages in four penumbral sectors.}
  \label{fig-methods}
  \end{center}
\end{figure}

The bulk of our analysis considers the time series of the magnetic field azimuth in the sunspot. Fig. \ref{fig-methods} shows the spatial locations where the azimuth variation is analysed: 16 individual pixels in the top row, averages over two circumferences in the middle row, and averages over four sectors of the penumbra in the bottom row.

\begin{figure*}
  \begin{center}
  \includegraphics[width=1.0\textwidth]{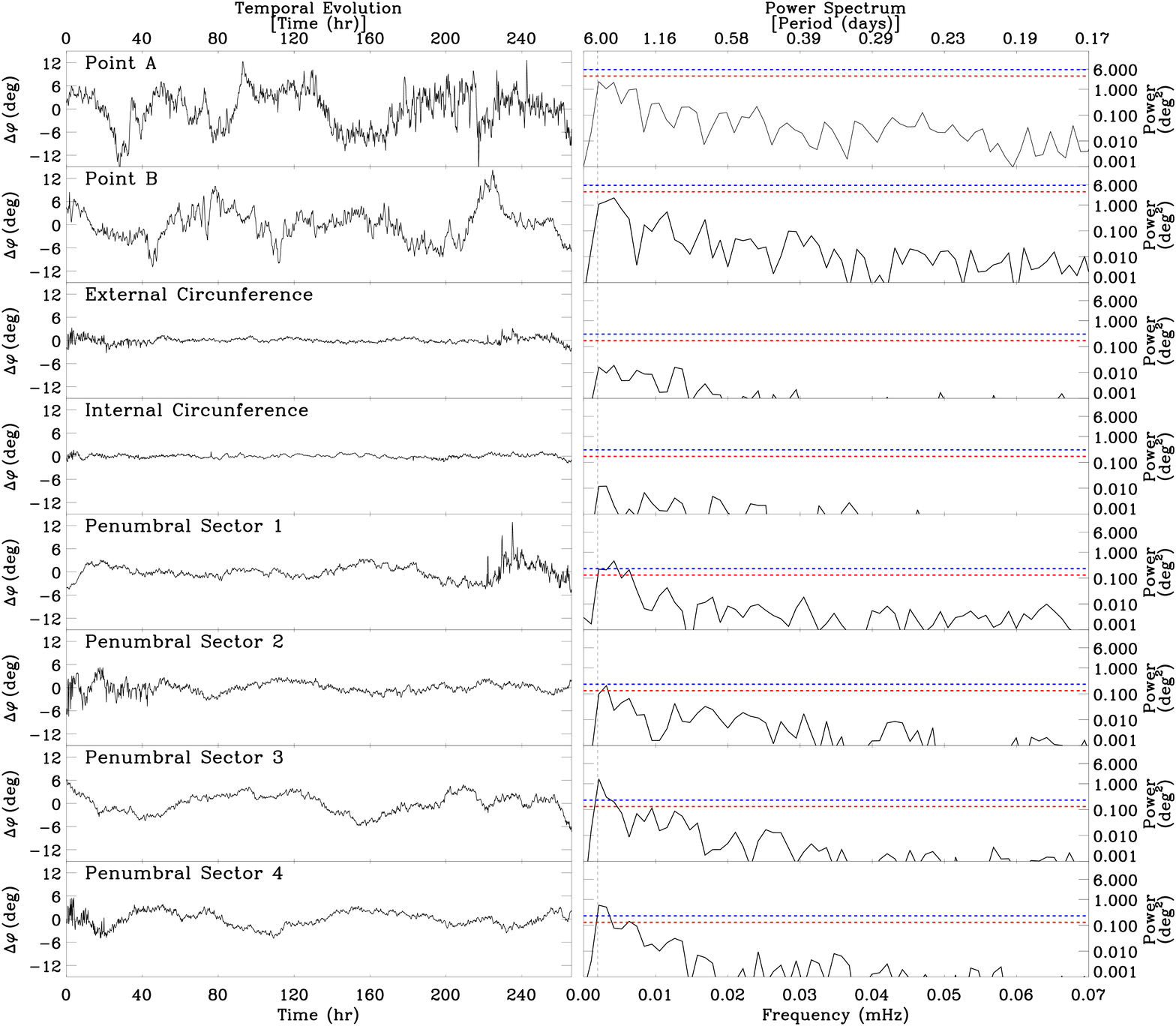}
  \caption{Left column: variation of the azimuth as a function of time (after the long-term trend was substracted) evaluated at: points A and B from Fig. \ref{fig-methods}, external and internal circumferences, and four penumbral sectors. Right Column: power spectra on logarithmic scale of the time series in the left panels. The vertical dashed line corresponds to the frequency associated to a six day period (our upper period limit) and the horizontal dashed lines are the confidence level for each case. The red one represents the $3\sigma$ level when the sunspot is close to the East limb and the blue one when it is crossing the central meridian.}
  \label{fig-resultsTogether}
  \end{center}
\end{figure*}

\begin{figure}
  \begin{center}
  \includegraphics[width=0.5\textwidth]{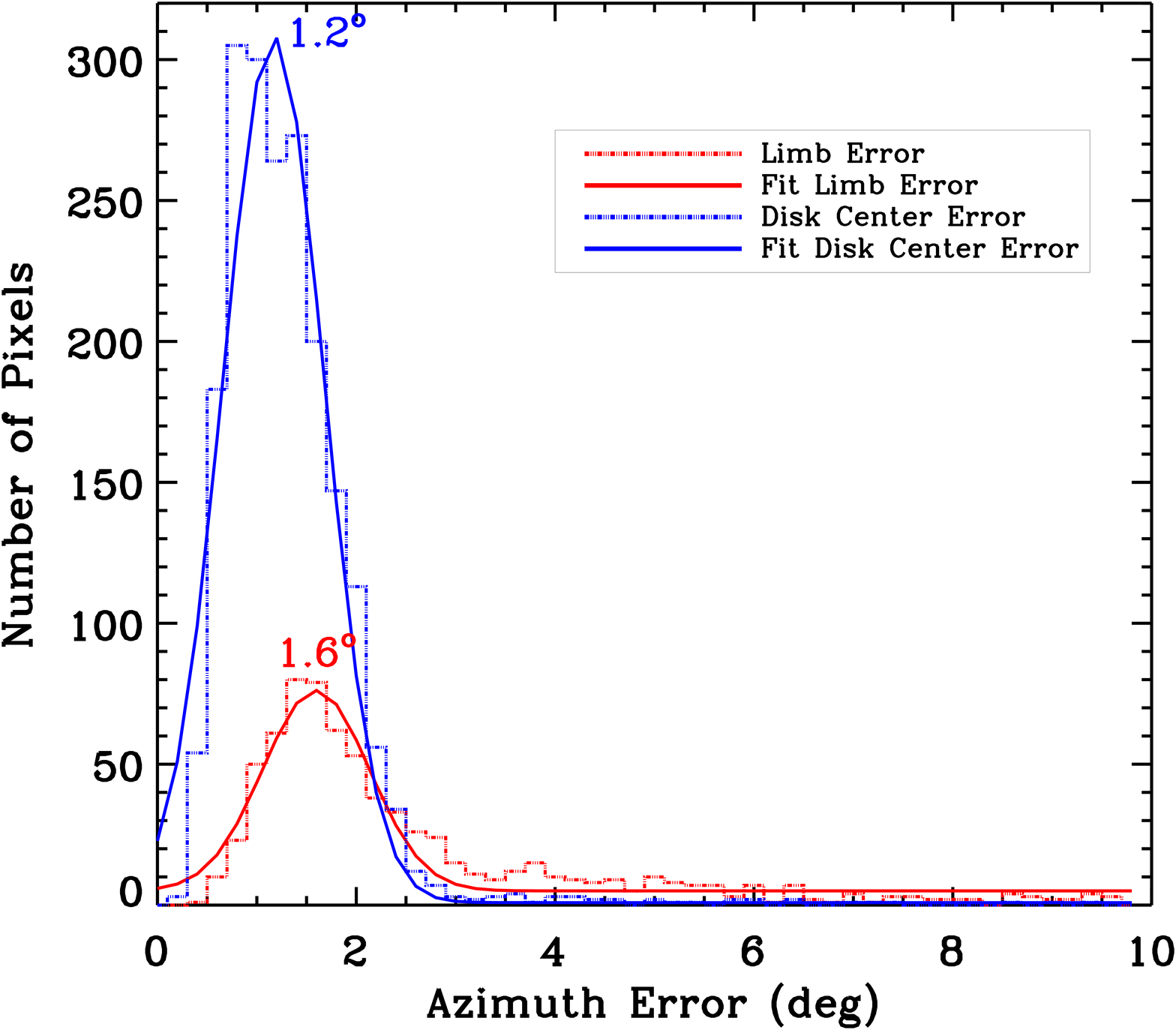}
  \caption{Dotted lines represent the azimuth error distribution of the sunspot penumbra for 2 different time steps: when the sunspot is at the East limb (red) and when the sunspot is crossing the central meridian (blue). Solid lines are the Gaussian fits to each distribution.}
  \label{fig-errors}
  \end{center}
\end{figure}

Figure \ref{fig-resultsTogether} shows the time-series plots (left) and their Fourier power spectra on logarithmic scale (right) of the magnetic field azimuth variation in the above-mentioned locations. The horizontal dashed lines in the power spectrum plots of Fig. \ref{fig-resultsTogether} represent three times the estimated noise level of the magnetic field azimuth at two different times: one when the sunspot is at the East limb (red lines; $3\sigma=4.8\deg$, which is $5.76\deg^2$ in the Fourier space), and the other when it crosses the central meridian (blue lines; $3\sigma=3.6\deg$, which is $3.24\deg^2$ in the Fourier space). In order to obtain the noise level, we made use of the azimuth error maps provided by the HMI team (\citealt{BSH+14}) for each frame. Then, we calculated the distribution of the errors in the penumbra of the sunspot for each one. These distributions and their Gaussian fits can be seen in Fig. \ref{fig-errors}. The location of the maxima was taken to be representative of the uncertainty value. We took these two frames because the error is smaller at disk centre and increases towards the limb. So any peak above the red/blue horizontal lines in Fig. \ref{fig-resultsTogether} would represent a detection of a torsional oscillation with an amplitude higher than the estimated confidence level. 

Also, we impose another criterion to consider an oscillation detection: there must be two complete periods at least. Then, we analyse the Fourier frequencies above 0.002 mHz (six days of period) since those offer sufficient sampling in timescales of days and below. The vertical dashed line in the power spectrum graphics denotes this frequency threshold.

The Fourier power spectrum at individual pixels (first and second rows of Fig. \ref{fig-resultsTogether}) does not show power clearly above the noise level. The noise is decreased to the point that we start to see some signal emerging slightly above the noise level in the low frequency range only when we average over may pixels (e.g. -Penumbral sector 1), but the amplitudes are extremely small, just some fractions of a degree. Such small amplitudes might be residuals from sunspot evolution or may even be due to the slight inaccuracies in the correlation tracking method. No signature of torsional oscillations, at the level defined in Table \ref{tab-previous_results} of the Introduction, is found in the analysis.

The time series at each pixel was constructed as the instantaneous value of the field azimuth minus its value in a reference frame. We chose the reference frame that corresponded to the passage of the spot through the central meridian. A global oscillation of the whole sunspot should exhibit its signature in the power spectrum, especially in the circular spatial averages. We have, however, found small spatial-scale transient oscillations in some penumbral filaments (Griñón-Marín et al. in preparation), yet no global oscillations are seen in these data.

\paragraph{Other sunspots.}
We repeated the same procedure for 24 other sunspots, spanning five years of cycle 24. All of these sunspots are mature alpha sunspots whose size does not change significantly while crossing the solar disk. 

\begin{figure*}
  \begin{center}
  \includegraphics[width=1.0\textwidth]{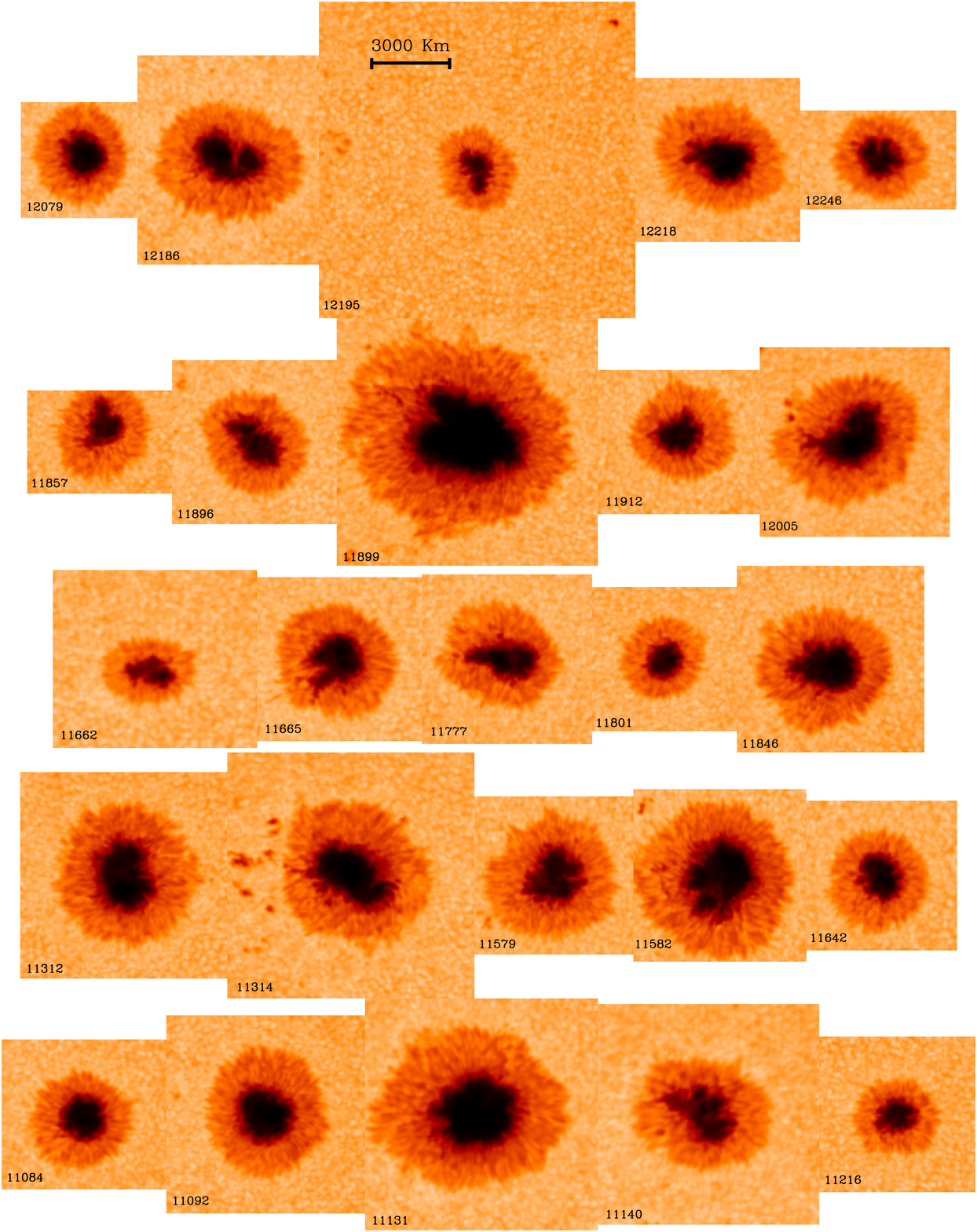}
  \caption{Images of all sunspots used in this work at the time of crossing the central meridian. The number on each picture corresponds to the NOAA active region number.}
  \label{fig-sunspots}
  \end{center}
\end{figure*}

Images of the sunspots analysed are shown in Fig. \ref{fig-sunspots}, with the same spatial and intensity scales, and their characteristics in Table \ref{tab-sunspots}: the NOAA active region number, the hemisphere where the sunspot was located, the start time, the duration, the size of the field of view, and the maximum amplitude of the azimuthal oscillation.

\begin{table*}
  \caption{Sunspots used in this work. The first column corresponds to the NOAA number of the active region associated with each sunspot. The second is the hemisphere where each sunspot was located. Third column is the time of the first frame used in our study. The time coverage (duration) is the fourth column, while the fifth column corresponds to the size of the field of view. Finally the last column represents the maximum oscillation amplitude, which was obtained by averaging the maximum amplitudes found with the circumference and the penumbral sector methods.}
  \label{tab-sunspots} 
  \centering
    \begin{tabular} {cccccc}
      \hline\hline
      Sunspot & Hemisphere & Start Time & Duration (d, h, m) & Size (Mm x Mm) & Max. Amp. ($\degr$) \\ 
      & & (yyyy.mm.dd$\_$hh:mm:ss UT) & & & \\
      \hline\hline
      11084 & South & 2010.06.26\_21:00:00 & 11, 03, 00 & 43.8 x 40.0 & 0.48 \\
      11092 & North & 2010.07.28\_18:00:00 & 11, 11, 48 & 52.9 x 52.9 & 0.37 \\
      11131 & North & 2010.12.02\_10:36:00 & 11, 14, 00 & 61.9 x 61.9 & 0.58 \\
      11140 & North & 2010.12.31\_10:00:00 & 11, 00, 00 & 58.9 x 58.9 & 0.52 \\
      11216 & South & 2011.05.16\_04:24:00 & 11, 19, 24 & 41.5 x 41.5 & 0.39 \\
      11312 & North & 2011.10.04\_15:00:00 & 12, 10, 24 & 55.1 x 55.1 & 0.58 \\
      11314 & North & 2011.10.11\_15:00:00 & 07, 10, 48 & 65.7 x 65.7 & 1.01 \\
      11579 & North & 2012.09.25\_09:48:00 & 09, 20, 36 & 42.3 x 42.3 & 0.69 \\
      11582 & South & 2012.09.27\_07:24:00 & 10, 02, 24 & 46.1 x 46.1 & 0.66 \\
      11642 & South & 2013.01.01\_03:00:00 & 09, 14, 24 & 40.0 x 40.0 & 0.56 \\
      11662 & North & 2013.01.23\_23:24:00 & 09, 14, 12 & 54.4 x 47.6 & 1.03 \\
      11665 & North & 2013.01.29\_07:00:00 & 10, 16, 48 & 43.8 x 43.8 & 0.86 \\
      11777 & South & 2013.06.20\_14:00:00 & 09, 04, 00 & 45.3 x 45.3 & 0.79 \\
      11801 & North & 2013.07.23\_13:36:00 & 09, 10, 12 & 38.5 x 38.5 & 0.84 \\
      11846 & South & 2013.09.17\_02:00:00 & 11, 13, 00 & 49.8 x 49.8 & 1.05 \\
      11857 & South & 2013.10.02\_10:00:00 & 09, 04, 48 & 38.5 x 27.5 & 0.97 \\
      11896 & North & 2013.11.10\_22:00:00 & 11, 09, 36 & 43.8 x 43.8 & 0.78 \\
      11899 & North & 2013.01.13\_09:00:00 & 10, 14, 48 & 69.5 x 66.1 & 0.59 \\
      11912 & South & 2013.12.02\_12:00:00 & 11, 04, 48 & 43.0 x 38.5 & 0.43 \\
      12005 & North & 2014.03.12\_13:00:00 & 11, 17, 48 & 50.6 x 50.6 & 0.77 \\
      12079 & North & 2014.06.01\_12:00:00 & 08, 22, 48 & 30.9 x 30.9 & 0.39 \\
      12186 & South & 2014.10.07\_18:36:00 & 11, 20, 12 & 48.3 x 55.9 & 0.53 \\
      12195 & North & 2014.10.24\_02:00:00 & 08, 10, 48 & 84.2 x 84.6 & 0.85 \\
      12218 & North & 2014.11.24\_22:00:00 & 10, 20, 24 & 43.8 x 43.8 & 0.48 \\
      12246 & North & 2014.12.22\_08:00:00 & 11, 12, 00 & 41.5 x 26.4 & 0.54 \\
      \hline
    \end{tabular}
\end{table*}

The results are consistent with those obtained for NOAA~11084. No torsional oscillations above the $3\sigma$ confidence level of the azimuth data were found.


\section{Conclusions}
\label{conclusions}
Using observations of the continuum intensity and vector magnetic field from SDO/HMI, we have sought evidence of torsional oscillations in 25 isolated sunspots. Since the HMI measurements are of sufficient sensitivity and time coverage, our analysis should be able to conclusively detect the kind of oscillations reported in previous observational (such as \citealt{G81}, \citealt{G85}, \citealt{GL87}, \citealt{HD03}, \citealt{GG05}, \citealt{G10}, and \citealt{GK11}) and theoretical (such as \citealt{G84}, \citealt{S84}, and \citealt{N97}) works. We emphasize that all previous works on this subject faced a formidable challenge, given the limited sampling and time coverage of the available observations. The SDO/HMI data are much more suitable for this kind of analysis, yielding more conclusive results. The SDO provides vector magnetic field measurements of the full solar disk every 12 minutes, allowing us to track the magnetic field azimuth with relatively high cadence and throughout the entire passage of the sunspots across the disk. We found no evidence of torsional oscillations in these 25 isolated sunspots. If such torsional oscillations exist, they are not a common occurrence in this type of sunspot.


\begin{acknowledgements}
The authors are grateful to the SDO/HMI team for their data. ABGM acknowledges Fundación La Caixa for the financial support received in the form of a Ph.D. contract. The National Center for Atmospheric Research (NCAR) is sponsored by the National Science Foundation. The authors gratefully acknowledge support from the Spanish Ministry of Economy and Competitivity through project AYA2014-60476-P (Solar Magnetometry in the Era of Large Solar Telescopes).
\end{acknowledgements}


\bibliographystyle{aa}
\bibliography{articulos} 

\end{document}